\documentclass[11pt,fleqn]{article}
\pdfoutput=1
\usepackage{amssymb,amsmath}
\usepackage{appendix}
\usepackage{graphicx}
\usepackage{caption}
\usepackage{color}		
\usepackage{epsfig}
\usepackage{float}

\newcommand{\be}{\begin{equation}}
\newcommand{\bea}{\begin{eqnarray}}
\newcommand{\eea}{\end{eqnarray}}
\newcommand{\ee}{\end{equation}}

\newcommand{\rs}{{\ensuremath{R_s}}}
\newcommand{\lp}{l_P}
\newcommand{\mv}{{\ensuremath{|0_M>}}}
\newcommand{\rv}{{\ensuremath{|0_R>}}}

\newcommand{\bh}{{\ensuremath{|BH>}}}
\newcommand{\mvbra}{{\ensuremath{<0_M|}}}
\newcommand{\rvbra}{{\ensuremath{<0_R|}}}

\newcommand{\bhbra}{{\ensuremath{<BH|}}}

\begin{document}
\begin{titlepage}
\bigskip

\bigskip\bigskip\bigskip\bigskip

\centerline{\Large \bf {Energy and Information Near Black Hole Horizons}}

\bigskip\bigskip
\bigskip\bigskip

\centerline{\bf Ben Freivogel\footnote{benfreivogel@gmail.com} }
\medskip
\centerline{\small ITFA and GRAPPA,
Universiteit van Amsterdam, Amsterdam, the Netherlands}

 \begin{abstract} 
The central challenge in trying to resolve the firewall paradox is to identify excitations in the near-horizon zone of a black hole that can carry information without injuring a freely falling observer. By analyzing the problem from the point of view of a freely falling observer, I arrive at a simple proposal for the degrees of freedom that carry information out of the black hole. An infalling observer experiences the information-carrying modes as ingoing, negative energy excitations of the quantum fields. In these states, freely falling observers who fall in from infinity do not encounter a firewall, but freely falling observers who begin their free fall from a location close to the horizon are ``frozen" by a flux of negative energy. When the black hole is ``mined," the number of information-carrying modes increases, increasing the negative energy flux in the infalling frame without violating the equivalence principle. Finally, I point out a loophole in recent arguments that an infalling observer must detect a violation of unitarity, effective field theory, or free infall.

\end{abstract}

\end{titlepage}

\section{Introduction}
The firewall paradox introduced by Almheiri, Marolf, Polchinski, and Sully (AMPS) \cite{amps} is a wonderful tool for sharpening our understanding  of black holes. I believe we will emerge from the firewall debate with either a shocking new quantum gravity effect or a sharper understanding of black hole complementarity.

The AMPS paradox is essentially a conflict between the following:
\begin{itemize}
\item To preserve unitarity, the outgoing Hawking radiation must carry information. There must be enough distinct allowed states of the quantum fields near the black hole to allow the radiation to carry information.
\item A freely falling observer naively expects to encounter a quantum state in the near-horizon zone that differs from the vacuum only at long wavelengths comparable to the Schwarzschild radius of the black hole.
\end{itemize}
So unitarity requires many allowed quantum states for the fields in the near-horizon zone, while free infall requires a state that is almost unique. In the remainder of the paper, I refer to the near-horizon region between the stretched horizon and $r \approx 3 G M$ simply as ``the zone."

Here I perform an elementary counting of the number of quantum states that are consistent with the naive expectations of an infalling observer. This is a simple quantum field theory analysis along the lines of the work of Unruh \cite{unruh} many years ago. I find that an infalling observer expects to encounter a flux of ingoing, negative energy excitations of the vacuum in the zone. These negative energy excitations are responsible for reducing the mass, and therefore the entropy, of the black hole, so arranging the correct quantum state for these excitations is sufficient for unitary evaporation of black holes. 

Doing the analysis in the infalling frame is convenient because this frame filters out the thermal noise that is present in the outside description and naturally picks out a subset of the zone degrees of freedom that may carry information.

\begin{figure}[hhh]
\begin{center}
\includegraphics[scale=.4]{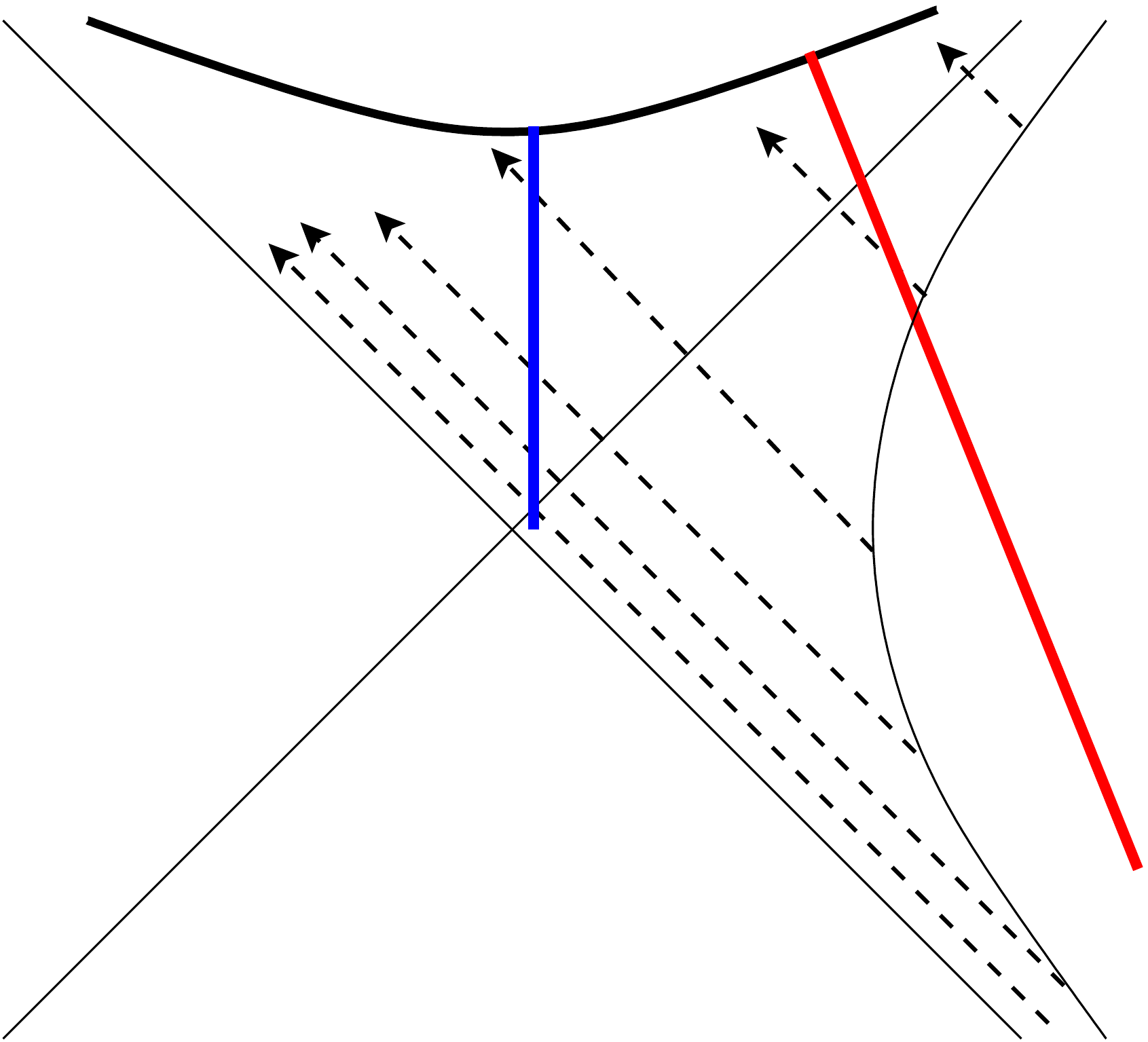}
\caption{From the infalling perspective, negative energy excitations are created around $r = 3 G M$ and fall into the black hole (dashed null rays). An observer falling in from infinity (solid red line) encounters a few of these excitations with wavelength of order \rs, but an observer who begins freely falling from near the horizon (solid blue line) will be frozen by collisions with blue-shifted negative energy excitations.}
\label{rindler}
\end{center}
\end{figure}


%

To roughly describe the states, every outgoing s-wave in the zone has about a 50\% chance of making it to infinity and a 50\% chance of reflecting off the geometry and returning to the black hole. (The probability is frequency-dependent, but for the frequencies that dominate Hawking radiation the probability of reflection is of order $1/2$.) It is perfectly consistent with the equivalence principle for any given quantum to be either reflected or transmitted. So the allowed states can be specified roughly by which s-waves reflect off the barrier. Roughly, if we restrict to frequencies of order the temperature, we can specify the state of each ingoing s-wave photon in the zone.

A natural question is how many distinct modes can be excited in the zone at a given time. The answer depends somewhat on the choice of time slicing, but I will show that slices of constant Schwarzschild time contain of order $\log S$ bits of information.

One might worry that these information-carrying excitations will violate the equivalence principle, in that freely falling observers will encounter high-energy quanta. In fact, I show that observers who begin their free fall from at least of order the Schwarzschild radius away from the horizon do not encounter high-energy quanta in these states. On the other hand, observers who somehow begin their free fall from a location very close to the horizon $do$ encounter high-energy excitations, although they are ``frozen" rather than burned. Such observers are highly boosted relative to the natural rest frame determined by the geometry, and the equivalence principle does not protect them. Therefore, this freezing is consistent with the equivalence principle, unlike the ``s-wave firewall" discussed in AMPS, which  affects even observers who fall in freely from infinity.

Negative energy excitations are possible in quantum field theory \cite{nege}, but they must be compensated by nearby regions of positive energy. In this case, the positive energy is behind the black hole horizon. This makes it clear that these ingoing excitations must be created in a manner that is delocalized over the entire zone.
So before worrying about delicate questions of unitarity, the description of the infalling observer is that the outgoing Hawking radiation is created in a manner that is delocalized over the entire zone. 

Very roughly, in the infalling description the region around $r = 3 GM$ sends positive energy particles out to infinity and negative energy excitations into the black hole.
To give some further intuition, from the outside point of view, I imagine that the stretched horizon is constantly having a delicate conversation with the potential barrier, carefully tuning itself so that the correct quanta are reflected off the barrier.
Another analogy is to think of the region between the stretched horizon and the barrier as a kind of electromagnetic cavity. The details of the boundary conditions at the stretched horizon and the barrier control the precise spectrum of electromagnetic excitations in the zone. From this point of view, we can think of the negative energy density in the zone as the result of the Casimir effect.

In order to use these states to avoid the firewall paradox, one must assume that most degrees of freedom in the zone of an old black hole are entangled with the stretched horizon rather than the early radiation. Consider high angular momentum modes in the zone. These modes are emitted from the stretched horizon, go a short distance into the zone, and then fall back in. It is natural to think that these modes are entangled with the stretched horizon. However, at late times the entire black hole plus zone system (suitably defined) has fewer degrees of freedom than the Hawking radiation that has already been emitted. Therefore, if the dynamics of the black hole are completely scrambling, then the black hole and the zone must separately be almost maximally entangled with the early radiation and not each other.

 Therefore, we require nontrivial dynamics of the stretched horizon- essentially it should contain special degrees of freedom that are ``bound" to the excitations in the zone and do not mix with the other degrees of freedom. This means that most modes in the zone are entangled with the stretched horizon, while the special quanta that are going to escape to infinity are entangled with the early radiation. While this is not the simplest possibility for the dynamics of the stretched horizon, a slight complication in a system we know nothing about is a small price to pay for preserving the equivalence principle. 

It is important to clarify which aspects of this paper are speculative, and which are not. The statement that certain freely falling observers encounter negative energy density is not speculative: it is true in effective field theory, and follows from the properties of the Unruh vacuum \cite{unruh}, as I describe in more detail in section 2. In addition, the estimates of how many states are consistent with the equivalence principle rely only on well-known physics. However, in effective field theory the choice of state for an evaporating black hole $is$ uniquely determined by the conditions that it is nonsingular at the future horizon and that no flux is incident from infinity. In order to access the different states consistent with the equivalence principle would require, for example, subtle modifications of the reflection coefficient for quantum fields reflecting off the barrier. I am not aware of any argument within effective field theory for such a modification. Therefore, as I describe in much more detail below, ideas about how the information gets into the ingoing quanta remain highly speculative.

 \paragraph{Conflict with effective field theory? ``Evolving back" $vs.$ ``evolving in."} One major problem remains: if we zoom in close to the horizon and focus on only the low angular momentum modes, there is a clean division between ingoing and outgoing waves. The states I described above contain information in the ingoing modes, while the outgoing modes are exactly thermal. This is perfectly consistent with unitarity of black hole evaporation until one asks $how$ the ingoing modes got into just the right state to preserve unitarity. The geometry outside the black hole should contain no information because of the no-hair theorem, so if the outgoing modes are exactly thermal then the ingoing modes cannot contain information. Someone has to put the information into the ingoing modes.

More concretely, if the outgoing Hawking quanta can be ``evolved back"  until they are localized very close to the horizon, as proposed by AMPS, then information must be stored in $outgoing$ quanta in the zone. But in the vacuum state,  outgoing quanta near the horizon should be nearly maximally entangled with their partners behind the horizon, so they cannot carry information. This is the firewall paradox of AMPS \cite{amps}.

 I present two related arguments pointing out what I believe is an important gap in the argument. First, if we literally evolve the  field operator corresponding to the outgoing Hawking radiation backward in time, it will evolve to an operator smeared over the entire backward lightcone (or at least its boundary), and not localized near the horizon.
 
  Instead of evolving back, we can try to ``evolve in."  In other words, instead of evolving the data backward in time, we can try to evolve it radially inward.
  Knowing the black-hole S-matrix roughly gives us all the information about the field modes far from the black hole. We can then try to use the data at large r to constrain the data near the black hole.  ``Evolving in" is somewhat familiar from AdS/CFT, but it is a much more delicate operation than standard time evolution. 
  
Even in classical field theories, the question of evolving data radially inward is nontrivial. I explain a simple, geometric criterion from the theory of partial differential equations to diagnose when the data can be evolved in. This criterion tells us that inward radial evolution breaks down precisely at the outer edge of the zone, $r = 3 G M$. The data at large $r$ can be evolved in all the way to the angular momentum barrier, but not into the zone. This is a strong motivation to consider more carefully an essential feature of the AMPS argument: evolving back the outgoing Hawking radiation in order to obtain information about the quantum fields near the horizon.
  
One can try to get around this obstacle by Fourier transforming on the sphere. Then, at least in the free field theory approximation, it $is$ possible to ``evolve in" mode by mode. However, we now encounter a different obstacle to stating the AMPS paradox: at least for a wide class of black holes, the behind-the-horizon partner of a mode with definite angular momentum does not fit into any single causal patch.  Therefore, we can escape from the firewall paradox by the same means we escaped from Hawking's information paradox: black hole complementarity \cite{comp} states that effective field theory can be trusted only within a single causal patch. (The geometry of the infalling patch will be described more fully in a forthcoming paper \cite{ishengandme}.)


\paragraph{Mining.} An objection to the idea that the zone can contain as few as $\log S$ information-carrying degrees of freedom is the mining argument. I show that the same analysis described above naturally leads to a larger number of allowed states in the zone when mining equipment is present. The mining equipment naturally changes the infalling vacuum on scales set by the nature of the mining; this again leads to just the right number of allowed states to preserve unitarity.

\paragraph{Relation to other work.} It is difficult to accurately place this work in the context of the flurry of activity following the firewall paper. The firewall argument has been clarified, strengthened, and generalized by \cite{ampss, mp, bousso, boussopure} among others.

This paper has some similarity to the ``balanced holography" proposal of the Verlindes \cite{balanced}. Part of the analysis, regarding what modes fit into a causal patch, is similar in spirit to the recent analysis of Ilgin and Yang \cite{isheng}. 
The work of Giddings and collaborators \cite{giddings} has some similarities, but here I try to preserve local, effective field theory in the causal patch.  W. Kim and collaborators \cite{kim} have also pointed out the negative energy experienced by certain freely falling observers near the horizon of an evaporating black hole. Mathur and Turton \cite{turton} also propose that certain observers in the zone may be injured, while observers who fall in from infinity are not; their proposal is in the context of fuzz balls.

 Papadodimas and Raju \cite{papadodimas} give an elegant description of how small nonlocalities can conspire to allow for a resolution of the firewall paradox; here I try to avoid introducing any nonlocality. Maldacena and Susskind \cite{erepr} have made an inspired proposal for when and how violations of effective field theory occur: old black holes develop an Einstein-Rosen bridge connecting them to their Hawking radiation, modifying the usual geometry behind the horizon; these and related ideas are also devoloped in \cite{susskind}. A different picture of the entanglement emerges in the analysis of Shenker and Stanford \cite{shenker}.  Harlow and Hayden \cite{hayden} have made a particularly interesting quantum computation argument against the firewall proposal. Many other proposals for escaping the paradox have been made, and the above is an unfair sampling.

Here, I do not propose any fabulous new effects in quantum gravity; rather, I suggest an escape from the firewall argument on its own terms.  I argue that the old, simple idea that any given observer can  have it all- unitarity, the equivalence principle, and effective field theory- is not yet ruled out.

\section{The Quantum State in the Zone}

\subsection{Which states are consistent with the equivalence principle?} We want to ask how unique the state of the quantum fields in the near-horizon zone is. How many degrees of freedom can be excited consistent with the naive expectations of an infalling observer?  Roughly, the equivalence principle dictates that a freely falling observer should detect of order one quantum with wavelength of order \rs\ as he falls through the zone.

For an eternal black hole, or a big, old black hole in AdS, the natural quantum state is the Hartle-Hawking state. Zooming in on the region near the horizon, the size of the 2-sphere is approximately constant, and the geometry looks like a 2-sphere cross 1+1 dimensional Rindler spacetime- that is, a 2-sphere cross 1+1 dimensional Minkowski space. The Hartle-Hawking state, when restricted to the zone, is roughly the Minkowski vacuum. This approximation becomes better and better as we zoom in closer to the horizon.

The quantum state near an evaporating black hole is not uniquely determined by the equivalence principle. A cartoon version is as follows. Focus on quanta with frequency set by the temperature of the black hole, $\omega \sim T \sim 1/\rs$. For these quanta, high angular momentum modes have an exponentially small chance of escaping to infinity, while s-waves have of order a 50\% chance of escaping. If we focus just on the s-waves, we can say that each outgoing quantum has a 50\% chance of escaping to infinity, and a 50\% chance of falling back into the black hole.\footnote{One might think that there is no angular momentum barrier for the s-wave, so it definitely makes it to infinity. However, in simply going from the $3+1$ action to the $1+1$ description, one finds a nontrivial radial potential even for the s-wave.}

Which particles get reflected back in is a delicate quantum process. A naive infalling observer would just say that any given quantum may or may not be reflected. He thus identifies a family of reasonable quantum states for the zone. A simple way to think about it is to say that every Schwarzschild time, $\Delta t \sim \rs$, 2 s-wave photons are emitted from the stretched horizon, one with each helicity. Depending on which photon escapes to infinity, we have a choice of the state for each of the ingoing photons, shown in figure \ref{rindler}. Another way to think about it is that tunneling through the barrier eliminates half of the ingoing photos that we would have had if the black hole were surrounded by a reflecting mirror. The choice of state in the zone is the choice of which ingoing quanta to erase.\footnote{This cartoon has the disadvantage that it might seem to depend crucially on the radiation being composed of particles with spin. An alternative cartoon that works even for scalar radiation is to say that the choice of state is a choice of which quantum field theory modes are populated by the outgoing radiation. I thank a referee for pointing out this shortcoming.} 

As mentioned in the introduction, there $is$ a natural choice of the quantum state for an evaporating black hole. Consider the usual Schwarzschild modes, with definite frequency with respect to the Schwarzschild time. Such modes see a potential barrier set by the black hole mass and the angular momentum of the mode in question. The natural orthonormal basis consists of 
\begin{itemize}
\item{1)} Modes incident from infinity, which are purely ingoing in the zone
\item{2)} Modes incident from the horizon, which are purely outgoing near infinity
\end{itemize}
The state of the modes of type 1 are determined by the requirement that there is no incoming flux from infinity, while the state of the modes of type 2 (which is actually a density matrix) is determined by the requirement that the future horizon is nonsingular. 
The statement in this section is simply that subtle deviations in the reflection coefficient with characteristic scale set by the Schwarzschild radius are consistent with the equivalence principle, and lead to the large number of allowed states described here.

\paragraph{Alternative characterization of states consistent with the equivalence principle.} Since it is confusing to identify states by which quanta are missing, consider another thought experiment that probes the uniqueness of the state of the quantum fields in the zone.  Suppose we want a black hole in asymptotically flat space not to evaporate. We can compensate the evaporation by throwing in quanta that are roughly similar to the outgoing Hawking radiation. In order that the expectations of an infalling observer, based on a naive application of the equivalence principle, are unaffected by the quanta we throw in, let us throw them in from infinity, roughly one quantum per Schwarzschild time with wavelength set by the size of the black hole. As long as we obey these rough guidelines, we will not be adding any new high-energy features to the black hole geometry.

But we can choose the detailed quantum state of the quanta we throw in. Since we are throwing in of order one quantum per light crossing time with wavelength of order the black hole radius, there is of order one bit of information per quantum. For example, one can send in photons and choose the helicity of each photon.  

Since the region near the horizon is well-described by Rindler space, it is convenient to translate the above description into a purely Rindler space description. In that case, we replace the asymptotic part of the black hole geometry by cutting off Rindler space at a distance \rs\ away from the horizon. Then the quanta thrown in from infinity become, in the Rindler description, quanta with wavelength \rs\ thrown in by a Rindler observer at a distance \rs\ away from the horizon. Figure \ref{rindler} can now be thought of as a picture of these real ingoing quanta (dotted null lines).

We can summarize the picture in a simple way. If we zoom in on a region deep in the zone, far from the angular momentum barrier, and focus on the s-waves, then we expect a unique state for the outgoing modes, but we have a choice of states for the ingoing modes, all of which are consistent with the equivalence principle.

There are two immediate questions.  Do these information-carrying modes in the zone injure an infalling observer? And is there enough information in them to allow for unitary black hole evaporation? I address these questions in the following two subsections.

\subsection{Experience of an infalling observer} 

It turns out that these excitations do not injure an observer who falls in from infinity, or anywhere outside the angular momentum barrier $r = 3 G M$. However, they do injure a freely falling observer who somehow arranges to begin freely falling from a location very close to the horizon. For example, consider an observer who is lowered by a rope and ``dangles" at constant Schwarzschild $r$ for some time. It clear that while he is still holding the rope, the observer will detect thermal radiation due to the Unruh effect. But even after letting go of the rope, this observer will encounter a large number of these ingoing quanta before hitting the singularity. 

This observer is in effect very boosted relative to the frame picked out by the geometry of the black hole. Roughly, quanta with wavelength of order \rs\ are emitted every \rs\ from a location \rs\ away from the horizon, but this observer sees them as high energy quanta because the ``source" is blue-shifted relative to him, as shown in figure \ref{rindler}.

The above description can be made more quantitative by examining the stress-energy tensor in the zone. At any point well away from the singularity, the stress tensor is small in the frame of an observer who freely falls from infinity. The stress tensor near a black hole was described by Unruh \cite{unruh} and others. There is one new ingredient here relative to Unruh's work. In Unruh's description, the outgoing s-waves are in the Minkowski vacuum, and therefore thermal from the Rindler point of view, while the ingoing s-waves are completely absent- they are in the Rindler vacuum.  The Rindler vacuum is again a unique state in the zone. In fact, as discussed above, some of the s-waves should reflect, so the ingoing s-waves should not be purely in the Rindler vacuum, but instead in a state that is a compromise between the Rindler and Minkowski vacua. 

We can be more quantitative. Computing the stress tensor in quantum field theory involves an ambiguity in regulating the infinities, but the difference in the stress tensor between two states is well-defined. If we go deep in the zone and treat only the s-wave, then the problem is $1+1$ dimensional Minkowski space.  Define coordinates so that
\be
ds^2 = dX^+ dX^-
\ee
Then for the simplest case, a free massless field, the stress tensor in the Minkowski vacuum $\mv$ differs from the stress tensor in the Rindler vacuum $\rv$ by:
\bea
\mvbra T_{++} \mv - \rvbra T_{++} \rv &\sim& {1 \over (X^+)^2} \\
\mvbra T_{--} \mv - \rvbra T_{--} \rv &\sim& {1 \over (X^-)^2 } \\
T_{+-} &=& 0
\eea
(Since the problem is 1+1 dimensional, the stress tensor has dimensions of mass squared.)

Now the black hole is like the Minkowski vacuum for the right movers, but it is missing half of the left movers, so deep in the zone we have
\bea
\bhbra T_{++} \bh - \mvbra T_{++} \mv &\approx& 0 \\
\bhbra T_{--} \bh - \mvbra T_{--} \mv &\sim& -{1 \over 2 (X^-)^2 }\\
T_{+-} &=& 0
\eea
A freely falling observer would naturally compare to the Minkowski vacuum, so the above equation is relevant for the experience of an infalling observer. Note the minus sign: a freely falling observer sees a $negative$ energy density relative to Minkowski spacetime. Such negative energy densities are known to be possible in quantum field theory \cite{nege} and have been constructed in simple examples. Isolated negative energy densities are not possible- a negative energy density must be compensated by a larger nearby positive energy density. (The field must ``pay back" for the negative energy density with a positive energy density, and it must pay interest on the negative energy it borrowed.) In this case, the compensating positive energy is behind the black hole horizon.

So before worrying about delicate questions of unitarity, it appears that the natural description of the infalling observer is that the outgoing modes are in the vacuum, but the ingoing modes are in one of a number of allowed states which carry negative energy into the black hole. Roughly, instead of thinking of particle creation at the stretched horizon, the infalling observer thinks of ``pairs" of particles produced in the angular momentum barrier: an ingoing negative energy particle, and an outgoing positive energy particle.  Since the negative energy particle is only possible due to compensating positive energy behind the horizon, the creation of the negative energy particle must be described as delocalized over the entire zone, but this is natural because it has wavelength of order \rs\ and is created about \rs\ from the horizon.  When we come to the question of unitarity, it is these pairs of particles that must be carefully managed in order to preserve unitarity in black hole evaporation.

Of course this is just a different description of the same physics seen by the external observer. The external observer naturally compares to the Rindler vacuum in the zone, and describes the same process in a more familiar way: the stretched horizon emits quanta, and some of them bounce off and come back into the black hole, while others escape to infinity. No quanta come in from infinity.

\subsection{Unitarity}
According to the description above, the states consistent with the equivalence principle consist, in the infalling description, of ingoing, negative energy excitations emitted from around the angular momentum barrier. Is the freedom in choosing the state of these excitations sufficient to allow for unitary evaporation? In fact it is: these excitations are precisely the thing that causes the black hole mass to decrease. Unitarity means that when old black holes lose mass they must also become less entangled with the environment. This can be arranged by carefully tuning the negative energy excitations.  In addition, the positive energy quanta emitted to infinity must also carry information.

Now a miracle is still necessary to get these modes into the correct state. I return later to the question of whether this miracle is possible within effective field theory.

The above argument for unitarity suffices, but to get a better feel for the information flow it is natural to count the states.  How many bits of information are there in the zone at one time? The answer depends somewhat on the choice of time slice; here, let's use the slices of constant Schwarzschild time natural for an outside observer.  

\begin{figure}[htbp]
\begin{center}
\includegraphics[scale=.5]{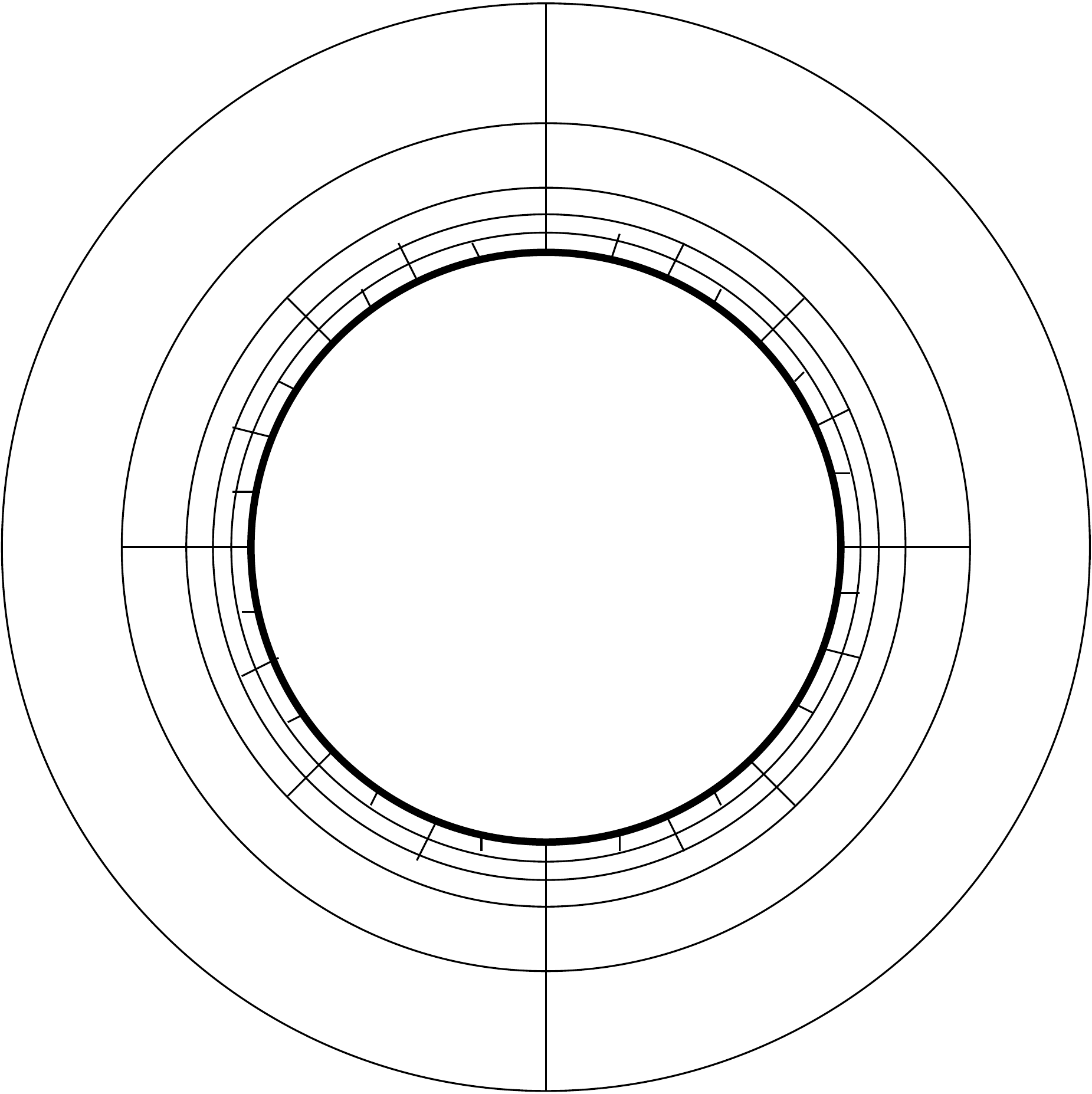}
\caption{From the outside point of view, the black hole atmosphere has of order one excited degree of freedom in each ``box" inside the zone. One can see more and more boxes near the horizon (thick line) as the higher angular momentum modes become excited. In the Hartle-Hawking state, the radiation contains no information, while an evaporating black hole contains one bit of information in each concentric ring.}
\label{zoneboxes}
\end{center}
\end{figure}

The evaporation process erases one photon per Schwarzschild time that was supposed to reflect back into the black hole. This erasing process happens at roughly the angular momentum barrier, $r \approx 3 G M$.  We would like to know how long this missing photon would have remained in the zone. 

The metric of a Schwarzschild black hole is
\be
ds^2 = - f(r) dt^2 + {dr^2 \over f(r)} dr^2 + r^2 d\Omega_2^2
\ee
with
\be
f(r) = 1 - {\rs \over r}~.
\ee
Ingoing photons satisfy
\be
dt = -{dr \over f(r)}
\ee
so the amount of Schwarzschild time $t$ that elapses between the time the photon would have been reflected off the barrier until it would have reached the stretched horizon can be calculated to be
\be
t \sim \rs \log {\rs \over \lp} \sim \rs \log S
\ee
where $S$ is the black hole entropy. This version of the result generalizes to many types of black holes in a variety of dimensions.

Since a single quantum takes of order $\log S$ units of Schwarzschild time to pass through the zone, and the black hole emits one quantum per Schwarzschild time, the zone must contain at least $\log S$ information-carrying quanta. 

 Therefore, it is consistent with the equivalence principle to expect that  $\log S$ bits of information are needed to characterize the state in the zone. 

%


\section{Conflict with effective field theory?}
The above analysis shows that, at the level of counting bits of information, there are enough states in the zone so that an infalling observer does not experience a firewall, while unitarity is preserved. But an important question remains.  If we focus on the s-waves in a region deep in the zone, the ingoing and outgoing waves do not interact. The discussion above involves encoding information in the ingoing waves, while the outgoing modes remain in a definite state. Having the correct state in the ingoing sector $is$ sufficient for black hole evaporation to be unitary, but how does the ingoing sector end up in exactly the correct quantum state to ensure unitarity? Is there a violation of effective field theory outside the black hole?

I believe this issue remains unresolved.   The basic issue is the following. We know that the radiation propagating out to infinity must contain information. For an old black hole, these outgoing Hawking quanta must be entangled with the early radiation, rather than with the remaining black hole. The issue is whether we can somehow ``evolve back" the outgoing Hawking quanta in order to obtain information about modes while they are still deep in the zone. 

Let me first describe an approximation where I believe it $is$ possible to evolve back the outgoing Hawking radiation until it is localized deep in the zone, leading to the firewall paradox. 
\begin{itemize}
\item Assume we can focus on the s-waves, and use the approximation that there is no mixing between ingoing and outgoing waves, and no mixing between modes of different angular momentum. In particular, this means that an outgoing s-wave near the horizon will certainly reach infinity.
\item Assume that the infalling observer can measure the s-wave behind the horizon-that is, the relevant Hawking partner of the outgoing radiation.
\end{itemize}
Given these assumptions, we $can$ evolve back, yielding the firewall paradox. (Actually, these assumptions are only sufficient for an ``s-wave firewall," but as AMPS have argued, even an s-wave firewall violates the equivalence principle. In this section I focus on the arguments for such an s-wave firewall, which are the strongest of all the firewall arguments.)

In this approximation, there is no mixing between ingoing and outgoing waves, so we can simply propagate the outgoing Hawking photons back along null rays until they are deep in the zone. 

\subsection{Evolving back $vs.$ evolving in}
However, in the real problem there is mixing between ingoing and outgoing modes: outgoing s-waves with frequencies of order the black hole temperature have an order one probability of reflecting off the geometry. (One might think that there is no angular momentum barrier for the s-wave, so it definitely makes it to infinity. However, in simply going from the $3+1$ action to the $1+1$ description, one finds a nontrivial radial potential even for the s-wave.) Therefore, literally evolving back the field operator corresponding to the outgoing mode yields an operator spread over the entire backward lightcone of the outgoing radiation, so it is spread over the entire zone and even extends beyond the zone.

Physically, however, this seems like a technicality. We know that no radiation is coming in from infinity, so we should be able to ignore at least the part of the operator that is outside the zone. Instead of doing a standard Cauchy evolution backward in time, the natural way to use the outgoing Hawking radiation to learn about the zone is to do ``radial evolution" inwards.  
This type of radial evolution is familiar from AdS/CFT, but in the presence of a black hole there is a surprising obstacle.

  We can address this question in the classical limit. Suppose we know the exact field configuration far from the black hole for all time, and we would like to reconstruct the solution near the horizon, but still outside the horizon. This is simply the classical limit of evolving back the Hawking radiation until it is near the horizon. If we ask the question in flat space, then there would be no problem in doing this reconstruction of the interior. Suppose that we know the solution everywhere outside a sphere of radius R for all time. Can we reconstruct the solution inside the sphere? In flat spacetime, the answer is yes.

More generally, however, radial evolution is a much more delicate problem than standard time evolution. This difference shows up already at the classical level: solving a known partial differential equation in a fixed background geometry. 

 Mathematicians \cite{math} have addressed this type of problem, called unique continuation: given the solution to a known differential equation in some region $r > r_*$ for all time, is there a unique continuation of the solution across the boundary $r = r_*$? There is a simple, intuitive diagnostic: if there are null geodesics that graze the surface $r = r_*$ but do not enter the known region $r > r_*$ where the solution is known, then the continuation is not unique. This was previously discussed in the AdS/CFT context in \cite{claire, stefanvlad}; see \cite{heemskerk} for related work.

This result is intuitive to physicists:  given a null geodesic, one can construct a solution that is localized arbitrarily close to the geodesic by going to a regime where the geometric optics approximation is good. If there are null geodesics that never enter the ``known" region $r > r_*$, then one can construct solutions that are arbitrarily close to 0 in the known region $r>r_*$, but are nonzero in the interior. In a sense, it may be the case that knowing the solution {\it exactly} in the region $r>r_*$ does determine the solution in the interior, but knowing the exterior solution to an arbitrary finite precision does not determine the interior solution to $any$ precision at all. This is usually summarized by saying that there is no {\it continuous mapping} between exterior data and interior data. It seems to me that continuity is the physically relevant criterion for whether reconstruction is possible.

This result is robust, and holds for a wide variety of differential equations. The essential feature is just that the high-frequency behavior of the equation is the same as the wave equation. I return in the next subsection to possible ways around the reconstruction theorem.

We can now analyze the behavior of null geodesics in the Schwarzschild solution. If we choose the affine parameterization, null geodesics are extrema of the action
\be
S = \int d\lambda g_{a b} \dot x^a \dot x^b
\ee
where dot denotes the derivative with respect to the affine parameter $\lambda$.

For metrics of the Schwarzschild form, and choosing the particle to move only in the $\theta$ direction on the $S_2$, this becomes
\be
S = \int d\lambda \left( f(r)  \dot t^2 - {\dot r^2 \over f(r)} - r^2\dot \theta^2 \right)
\ee
There are two conserved quantities, $E = f(r) \dot t$ and $\ell = r^2 \dot \theta$. Plugging these into the equation $ds^2 = 0$ for null curves gives
\be
- f(r) \dot t^2 + {\dot r^2 \over f(r)} + r^2 \dot \theta^2 = -{E^2 \over f(r)} + {\dot r^2 \over f(r)}   + {\ell^2 \over r^2} = 0
\ee
In other words, the radial motion of the null geodesics can be read off from an effective potential,
\be
\dot r^2 + V(r) = E^2 \ \ \ \ \ \ {\rm with} \ V(r) = \ell^2 {f(r) \over r^2}
\ee
For a Schwarzschild black hole, the function $f(r)$ is
\be
f(r) = 1 - {\rs \over r}
\ee
so the effective potential for null geodesics is
\be
V(r) = \ell^2 \left( {1 \over r^2} - {r_s \over r^3} \right)
\ee
This function approaches 0 at the horizon and at infinity, and has a maximum at $r = 3 r_s/2$, or in other words at the location of the unstable circular null orbit, $r = 3 G M$. Therefore, for any value of the angular momentum $\ell$, geodesics with small enough energy $E$ can bounce off the angular momentum barrier and never enter the known region $r > 3 G M$. 

Therefore, according to the theorem, the solution at large $r$ can be uniquely evolved radially inward until we reach the peak of the angular momentum barrier at $r = 3GM $.  
So classically evolving Hawking photons inward fails for the near horizon region $r < 3 G M$. (I return later to possible ways around the theorem, such as restricting to the s-wave.)

The reason reconstruction fails is physically clear, and does not require mathematical theorems. Due to the angular momentum barrier, photons are constantly emitted by the stretched horizon, bounce off the angular momentum barrier, and fall back into the black hole. Such photons can interfere constructively or destructively with with photons that do manage to escape the barrier. The existence of photons that are bound inside the angular momentum barrier means that knowing the state of the outgoing Hawking quanta is not enough information to reconstruct the field near the horizon. 

\subsection{Ways around the no-reconstruction theorem?} Despite the theorem, intuitively one should be able to learn $something$ about the solution inside $r=3GM$. The various arguments for the firewall paradox including \cite{amps, ampss, bousso} all rely on some technique of evolving back the outgoing Hawking radiation to obtain information about the degrees of freedom in the zone. 

For example, let us return to the approximation stated at the beginning of this section:  simply Fourier transform on the sphere, and evolve radially inward mode by mode.
Essentially, this approximation gets around the theorem by using the fact that in free field theory, modes with different angular momentum do not mix, so for any definite angular momentum we have a 1+1 dimensional problem. Suppose we focus on just one angular mode, say the s-wave.  In a 1+1 problem we can reverse space and time at the price of changing positive mass squared to negative mass squared. However, even for fields with negative mass squared, the time evolution for a finite time is well-defined. Therefore, the radial evolution across $r=3GM$ must be possible, mode by mode, in a free field theory.

 One obvious objection is that the fields we are dealing with must interact at least gravitationally, so in fact we cannot treat the angular momentum modes separately. Quantitatively, however, it may be possible to argue that these interactions are small enough that they can be neglected. If we work in terms of angular modes, we can perturb around the free field case, and it appears that no problems will arise at any order in perturbation theory \cite{heemskerk}. However, we do encounter problems with trying to reconstruct any operator with finite support in position space; see \cite{claire, stefanvlad} for a much more detailed discussion.

\paragraph{Angular modes inside the horizon may not fit in one causal patch.} My favorite objection pertains to the geometry of the causal patch. Suppose that we $do$ restrict to the s-wave, and that we do succeed in evolving it back until it is close to the horizon. Now we appear to have a problem: on one hand, this near-horizon s-wave must be entangled with the early radiation, but on the other hand it should be entangled with its partner behind the horizon.

But the partner of the s-wave is roughly the s-wave behind the horizon, or in any case it is some mode that is spread over the entire 2-sphere. But then we must ask whether this behind-the-horizon partner mode fits inside the causal patch of the infalling observer. There is only a conflict with black hole complementarity if there is a conflict with effective field theory within a single causal patch.

The geometry of the causal patch behind the horizon will be described more completely in a separate publication \cite{ishengandme}. For Schwarzschild and Reissner-Nordstrom black holes in 3+1 and higher dimensions, the infalling patch does not contain the entire horizon sphere. Therefore, for these black holes, complementarity offers an escape from the firewall paradox.
Note that this objection alone is sufficient to escape from the s-wave AMPS argument, without any need for the discussion in the remainder of the paper.

Of course, one can build a mode that is {\it approximately} localized in a finite angular region on the sphere out of relatively low $l$ modes, so it may seem like a mere technicality that we do not have access to the entire sphere. However, the fact that the reconstruction problem is impossible for any operator that is $exactly$ localized in a finite region motivates careful consideration of this technicality.

 For Kerr black holes the question is more subtle because there is no spherical symmetry, and we are still investigating whether the relevant partner mode fits inside the infalling causal patch. Geometrically, there are some 2-spheres that $do$ fit inside a causal patch, so it is possible that the firewall argument will work in this case, but further analysis is needed.

\section{Mining}
One urgent question about the plausibility of this proposal has to do with all the other modes in the zone. Suppose we accept for now that for the s-waves there are enough states in the zone to allow for unitarity evaporation, as I have argued above. What now about the other modes in the zone? I am claiming many degrees of freedom in the zone remain entangled with the black hole, rather than the early radiation.

This is dangerous because there is a technique for allowing additional modes to escape to infinity called black hole mining. The original mining procedure described by Unruh and Wald \cite{unruhwald} involved lowering boxes into the zone to capture the Hawking radiation. Recently, A. Brown \cite{adam} showed that in fact any box that obeys the null energy condition is worse than superfluous, and the optimal mining procedure is simply to lower a string into the zone. 

\begin{figure}[htbp]
\begin{center}
\includegraphics[scale=.6]{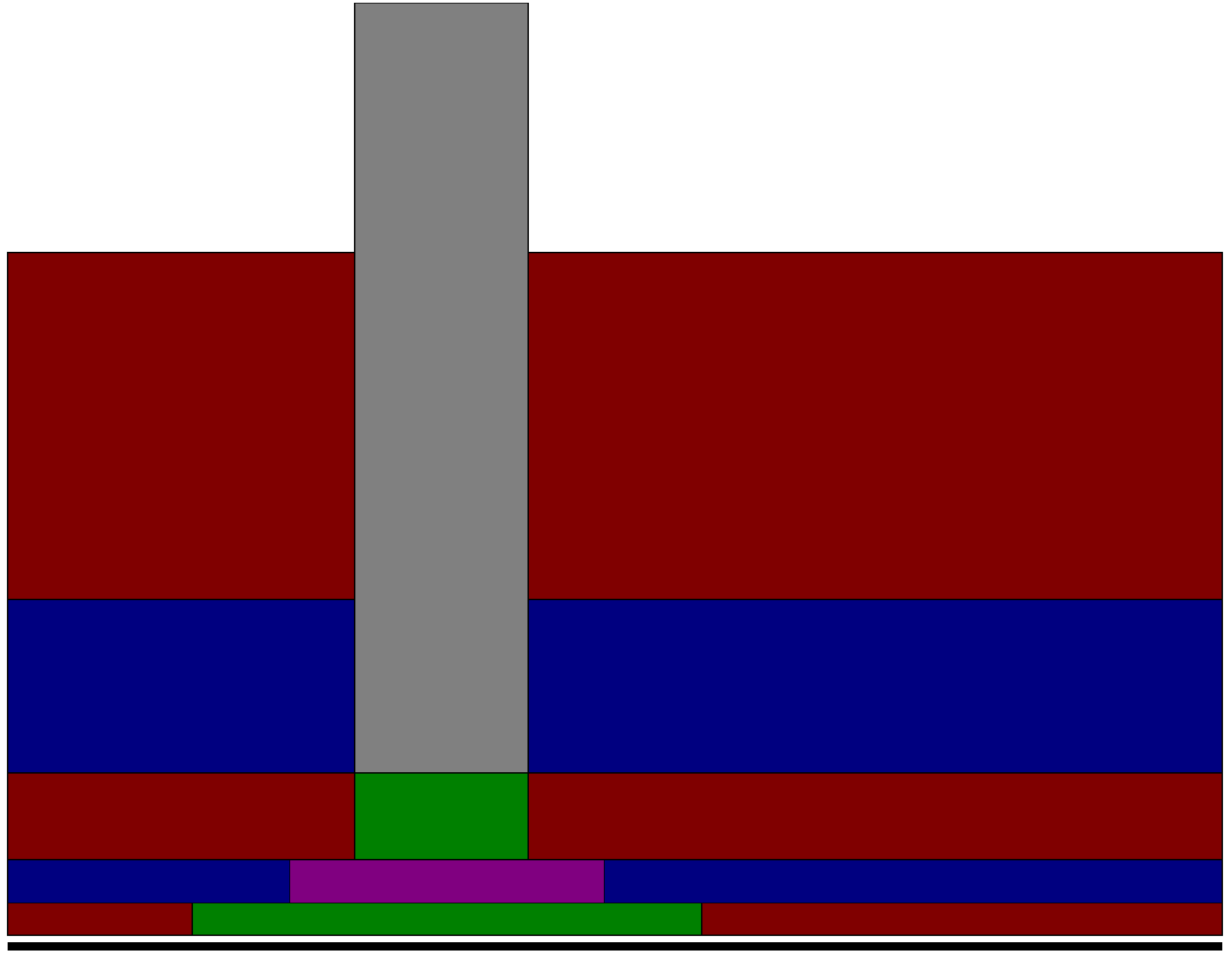}
\caption{The presence of a mining string (gray) in the zone changes the quantum state. From the infalling perspective, the string emits negative energy excitations that propagate into the black hole (green and purple boxes). Even in the absence of the mining equipment, evaporation effectively produces ingoing, negative energy s-waves (red and blue boxes). The alternating colors indicate the freedom in choosing the quantum state- each missing photon can be either left-handed or right-handed. The string is ``fat" because the maximum distance it can reach into the zone is related to the characteristic energy scale of the string, which also controls its width. }
\label{mining}
\end{center}
\end{figure}

The string cannot be lowered all the way to the stretched horizon without breaking, but it can be lowered far inside the angular momentum barrier. The tension of the string controls how far it can be lowered without breaking; see figure \ref{mining}. The string should couple to the field that we want to extract from the black hole. Then the quantum of interest can be emitted by the black hole, and instead of falling back in due to the angular momentum barrier, it can be absorbed by the string. Once it has been absorbed, it can then travel up the string- simply a 1+1 dimensional problem with no angular momentum barrier. Therefore lowering strings into the horizon allows us to extract high-l quanta from the black hole. The process of emitting a high-l quantum to infinity with the help of a string decreases the mass and entropy of the black hole. Therefore unitarity requires that the  quanta absorbed by the string must be entangled with the early radiation and not the black hole horizon. 

In order to preserve unitarity, it must be the case that lowering a string into the zone $changes$ the quantum state in such a way that the modes absorbed by the string carry information.  In the presence of mining equipment, the zone must carry not only the $\log S$ bits of information described above, but an additional number of bits that will be absorbed by the mining equipment.

 As Brown showed, a string with a given tension can be lowered only to a distance above the horizon related to its tension. If $\mu$ is the tension of the string, then it can be lowered only to the location where the local Hawking temperature is related to the tension $\mu = T^2$. Therefore, the proper distance $x$ from the horizon is given by
\be
x^2 = {1 \over \mu}
\ee
 In order for the mining to be effective- for the mining equipment to extract more energy than it emits- the mining apparatus cannot be moved quickly into place, nor can it be quickly turned on and off. As a result, the important question is essentially a static one: how does the quantum state of the black hole differ in the presence of a string designed to mine the black hole?

Turning again to the perspective of the infalling observer,  by the same argument as above the string will effectively emit negative energy particles with a wavelength set by the location of the endpoint of the string into the black hole. From the outside point of view, the string is absorbing some of the quanta of the field, dominantly at the endpoint of the string. But for the string to be effective at mining, it must be ``cold" and not send quanta into the black hole from infinity. Just as above, the black hole state now differs from the Minkowski vacuum near the horizon due to the missing ingoing quanta that should have been reflected by the angular momentum barrier but instead were absorbed by the string.  The black hole is also missing the s-waves that escaped, as before.

These additional negative energy excitations cause the black hole to evaporate more quickly. For the same reason as above, the quantum state of these negative energy excitations must be correct in order for unitary evaporation. Infalling observers will notice some additional structure in the quantum state near the mining string. This additional structure may injure some observers, but it does not violate the equivalence principle. 

The discussion of whether there is a conflict with effective field theory also mirrors the s-wave construction. The same theorem prohibits evolving radially inward; one can again try to avoid the theorem by decomposing in angular modes, but these angular modes still do not fit in any causal patch for Schwarzschild or Reissner-Nordstrom black holes.

To summarize, the presence of mining equipment in the zone (say a string) naturally changes the state seen by the infalling observer, with structure on the scale set by the location of the string. The new state still contains enough information-carrying modes in order to allow for unitarity without violating the equivalence principle.

\section{Conclusions and future directions}
I have presented what I consider an attractive picture of which modes in the near-horizon zone carry information. In this description, even for an old black hole, most modes in the zone remain entangled with the stretched horizon rather than the early radiation.

Note that the arguments here still leave a lot of freedom about precisely how many modes in the near-horizon zone are $not$ entangled with the black hole. In order to allow for unitary evaporation, at least of order $\log S_{\rm BH}$ modes must be entangled with the early radiation or in a pure state by themselves \cite{boussopure}, and not entangled with the  black hole.

I have pointed out what I believe is a significant loophole in the arguments that the infalling observer must detect a violation of unitarity, effective field theory, or free infall.
There are two main ingredients in this loophole. First, theorems from the theory of partial differential equations strongly suggest an obstacle to evolving the S-matrix data radially inward into the near-horizon zone. This obstacle can be avoided by working in terms of angular momentum modes on the 2-sphere, but this brings us to a second obstacle: at least for the simplest black holes, the angular momentum modes do not fit inside a single causal patch. Therefore the old idea of black hole complementarity, that no single observer will see a violation of effective field theory, has not been ruled out. 

This loophole essentially allows some number of outgoing modes in the zone to carry information, because their partners behind the horizon do not fit into any single causal patch. As stated, it applies to any mode that is delocalized over the entire sphere- far more than the $\log S_{\rm BH}$ modes necessary for unitarity. So these arguments do not yet fix the number of information-carrying degrees of freedom in the zone.

It is of course reasonable for effective field theory to be violated by a small amount even within one causal patch- the challenge is to identify which observables receive significant corrections.  However, the simplest possibility remains that effective field theory is not corrected in any significant way for causal patches with low curvature.  

Fortunately, the issues I have raised can be addressed using known techniques. We can analyze the geometry of the infalling causal patch in combination with a field theory analysis of which modes should be entangled  for a variety of black holes. Some results in this direction will appear in \cite{ishengandme}. In addition, there may be other ways of tightening the arguments in favor of the firewall paradox. For example, in spite of the motivating classical theorems, perhaps there is a clear argument that in quantum field theory the outgoing Hawking radiation can be evolved back to constrain the fields near the horizon.

I have not addressed the AdS/CFT version of the firewall argument \cite{mp} here. However, the obstacles to evolving radially inward discussed here appear in a very similar form in that context \cite{claire, stefanvlad}.

\section*{Acknowledgements}
I would like to thank Jos\'e Barb\'on, Jan de Boer, Raphael Bousso, Adam Brown, Bartek Czech, Dan Harlow, Irfin Ilgin, Ted Jacobson, Robert Jefferson, Laurens Kabir, Javier Mart\'inez Mag\'an, Stephen Shenker, Douglas Stanford,  Lenny Susskind,  Erik Verlinde,  and Alex Westphal for interesting discussions. I particularly thank I-Sheng Yang for many discussions and collaboration. I thank the organizers and participants of the  KITP workshop ``Fuzz or Fire" in August 2013 for including me in a stimulating workshop.


\begin{thebibliography}{999}


\bibitem{amps} 
  A.~Almheiri, D.~Marolf, J.~Polchinski and J.~Sully,
  ``Black Holes: Complementarity or Firewalls?,''
  JHEP {\bf 1302}, 062 (2013)
  [arXiv:1207.3123 [hep-th]].
  
\bibitem{unruh} 
  W.~G.~Unruh,
  ``Notes on black hole evaporation,''
  Phys.\ Rev.\ D {\bf 14}, 870 (1976).
  
\bibitem{nege} 
  L.~H.~Ford,
  ``Negative Energy Densities in Quantum Field Theory,''
  Int.\ J.\ Mod.\ Phys.\ A {\bf 25}, 2355 (2010)
  [arXiv:0911.3597 [quant-ph]].
  
  E.~E.~Flanagan,
  ``Quantum inequalities in two-dimensional curved space-times,''
  Phys.\ Rev.\ D {\bf 66}, 104007 (2002)
  [gr-qc/0208066].

\bibitem{comp} 
  L.~Susskind, L.~Thorlacius and J.~Uglum,
  ``The Stretched horizon and black hole complementarity,''
  Phys.\ Rev.\ D {\bf 48}, 3743 (1993)
  [hep-th/9306069].
  
  L.~Susskind and L.~Thorlacius,
  ``Gedanken experiments involving black holes,''
  Phys.\ Rev.\ D {\bf 49}, 966 (1994)
  [hep-th/9308100].
  
  L.~Susskind,
  ``The World as a hologram,''
  J.\ Math.\ Phys.\  {\bf 36}, 6377 (1995)
  [hep-th/9409089].
  
  \bibitem{ishengandme}
  B.~Freivogel and I-S.~Yang, {\it in preparation}.

\bibitem{ampss} 
  A.~Almheiri, D.~Marolf, J.~Polchinski, D.~Stanford and J.~Sully,
  ``An Apologia for Firewalls,''
  JHEP {\bf 1309}, 018 (2013)
  [arXiv:1304.6483 [hep-th]].

\bibitem{mp} 
  D.~Marolf and J.~Polchinski,
  ``Gauge/Gravity Duality and the Black Hole Interior,''
  Phys.\ Rev.\ Lett.\  {\bf 111}, 171301 (2013)
  [arXiv:1307.4706 [hep-th]].

\bibitem{bousso} 
  R.~Bousso,
  ``Complementarity Is Not Enough,''
  Phys.\ Rev.\ D {\bf 87}, 124023 (2013)
  [arXiv:1207.5192 [hep-th]].


  R.~Bousso,
  ``Frozen Vacuum,''
  arXiv:1308.3697 [hep-th].
  
\bibitem{boussopure} 
  R.~Bousso,
  ``Firewalls From Double Purity,''
  Phys.\ Rev.\ D {\bf 88}, 084035 (2013)
  [arXiv:1308.2665 [hep-th]].




  
\bibitem{balanced} 
  E.~Verlinde and H.~Verlinde,
  ``Passing through the Firewall,''
  arXiv:1306.0515 [hep-th].

\bibitem{isheng} 
  I.~Ilgin and I-S.~Yang,
  ``Causal Patch Complementarity: The Inside Story for Old Black Holes,''
  arXiv:1311.1219 [hep-th].
  
\bibitem{hayden} 
  D.~Harlow and P.~Hayden,
  ``Quantum Computation vs. Firewalls,''
  JHEP {\bf 1306}, 085 (2013)
  [arXiv:1301.4504 [hep-th]].
  
\bibitem{giddings} 
  S.~B.~Giddings and Y.~Shi,
  ``Effective field theory models for nonviolent information transfer from black holes,''
  arXiv:1310.5700 [hep-th].

  S.~B.~Giddings,
  ``Statistical physics of black holes as quantum-mechanical systems,''
  Phys.\ Rev.\ D {\bf 88}, 104013 (2013)
  [arXiv:1308.3488 [hep-th]].

  S.~B.~Giddings,
  ``Nonviolent information transfer from black holes: a field theory parameterization,''
  Phys.\ Rev.\ D {\bf 88}, 024018 (2013)
  [arXiv:1302.2613 [hep-th]].

  S.~B.~Giddings,
  ``Nonviolent nonlocality,''
  Phys.\ Rev.\ D {\bf 88}, 064023 (2013)
  [arXiv:1211.7070 [hep-th]].
  
  \bibitem{turton}
  S.~D.~Mathur and D.~Turton,
  JHEP {\bf 1401}, 034 (2014)
  [arXiv:1208.2005 [hep-th]].
  
  S.~D.~Mathur and D.~Turton,
  ``The flaw in the firewall argument,''
  Nucl.\ Phys.\ B {\bf 884}, 566 (2014)
  [arXiv:1306.5488 [hep-th]].
  
  \bibitem{kim}
  W.~Kim and E.~J.~Son,
  ``Freely Falling Observer and Black Hole Radiation,''
  Mod.\ Phys.\ Lett.\ A {\bf 29}, 1450052 (2014)
  [arXiv:1310.1458 [hep-th]].
  
  M.~Eune, Y.~Gim and W.~Kim,
  ``Something special at the event horizon,''
  arXiv:1401.3501 [hep-th].

\bibitem{papadodimas}
  K.~Papadodimas and S.~Raju,
  ``The unreasonable effectiveness of exponentially suppressed corrections in preserving information,''
  Int.\ J.\ Mod.\ Phys.\ D {\bf 22}, 1342030 (2013).

  K.~Papadodimas and S.~Raju,
  ``State-Dependent Bulk-Boundary Maps and Black Hole Complementarity,''
  arXiv:1310.6335 [hep-th].

  K.~Papadodimas and S.~Raju,
  ``The Black Hole Interior in AdS/CFT and the Information Paradox,''
  arXiv:1310.6334 [hep-th].

  K.~Papadodimas and S.~Raju,
  ``An Infalling Observer in AdS/CFT,''
  JHEP {\bf 1310}, 212 (2013)
  [arXiv:1211.6767 [hep-th]].
  
\bibitem{erepr} 
  J.~Maldacena and L.~Susskind,
  ``Cool horizons for entangled black holes,''
  arXiv:1306.0533 [hep-th].
  
  \bibitem{susskind}
  L.~Susskind,
  ``Butterflies on the Stretched Horizon,''
  arXiv:1311.7379 [hep-th].

  L.~Susskind,
  ``New Concepts for Old Black Holes,''
  arXiv:1311.3335 [hep-th].

  L.~Susskind,
  ``Black Hole Complementarity and the Harlow-Hayden Conjecture,''
  arXiv:1301.4505 [hep-th].

  L.~Susskind,
  ``The Transfer of Entanglement: The Case for Firewalls,''
  arXiv:1210.2098 [hep-th].

  L.~Susskind,
  ``Singularities, Firewalls, and Complementarity,''
  arXiv:1208.3445 [hep-th].
  
\bibitem{shenker} 
  S.~H.~Shenker and D.~Stanford,
  ``Multiple Shocks,''
  arXiv:1312.3296 [hep-th].
\bibitem{Shenker:2013pqa} 
  S.~H.~Shenker and D.~Stanford,
  ``Black holes and the butterfly effect,''
  arXiv:1306.0622 [hep-th].
  
  
  
  
\bibitem{unruhwald} 
  W.~G.~Unruh and R.~M.~Wald,
  ``Acceleration Radiation and Generalized Second Law of Thermodynamics,''
  Phys.\ Rev.\ D {\bf 25}, 942 (1982).
  
  W. G. Unruh and R. M. Wald, ``How to mine energy from a black hole," Gen.
Rel. Grav. D {\bf15}, 195 (1983).

\bibitem{adam} 
  A.~R.~Brown,
  ``Tensile Strength and the Mining of Black Holes,''
  Phys.\ Rev.\ Lett.\  {\bf 111}, 211301 (2013)
  [arXiv:1207.3342 [gr-qc]].


  
    
  
\bibitem{math}
  D. Tataru. Unique continuation problems for partial differential equations - Springer. in Geometric Methods in Inverse Problems and PDE Control, vol. 137 of The IMA Volumes in Mathematics and its Applications 2004.
  
\bibitem{claire} 
  R.~Bousso, B.~Freivogel, S.~Leichenauer, V.~Rosenhaus and C.~Zukowski,
  ``Null Geodesics, Local CFT Operators and AdS/CFT for Subregions,''
  Phys.\ Rev.\ D {\bf 88}, 064057 (2013)
  [arXiv:1209.4641 [hep-th]].
  
\bibitem{stefanvlad} 
  S.~Leichenauer and V.~Rosenhaus,
  ``AdS black holes, the bulk-boundary dictionary, and smearing functions,''
  Phys.\ Rev.\ D {\bf 88}, 026003 (2013)
  [arXiv:1304.6821 [hep-th]].
  
\bibitem{heemskerk} 
  I.~Heemskerk, D.~Marolf, J.~Polchinski and J.~Sully,
  ``Bulk and Transhorizon Measurements in AdS/CFT,''
  JHEP {\bf 1210}, 165 (2012)
  [arXiv:1201.3664 [hep-th]].
  
  
  
\end{thebibliography}
\end{document}